\documentclass[%
reprint,
superscriptaddress,
nofootinbib,
 amsmath,amssymb,
 aps,
 prl,
 longbibliography
]{revtex4-2}

\usepackage{hyperref}
\usepackage{subcaption}
\usepackage{caption}
\usepackage{graphicx}
\usepackage{graphicx}
\usepackage{dcolumn}
\usepackage{bm}
\usepackage{cleveref}
\usepackage{microtype}
\usepackage{amsmath}
\usepackage{graphicx}

\newcommand{\aveomega}{\left\langle \omega \right\rangle}
\newcommand{\pomega}{P\left(\omega\right)}
\newcommand{\cH}{$\mathcal{H}$}
\begin{document}

\title{Recognition Capabilities of a Hopfield Model with Auxiliary Hidden Neurons}

\author{Marco Benedetti}
\affiliation{Universit\`a di Roma La Sapienza, Rome, Italy} 
\author{Victor Dotsenko}
\affiliation{Sorbonne Université, CNRS, Laboratoire de Physique Théorique de la Matière Condensée (UMR 7600), 4 Place Jussieu, F-75252 Paris
Cedex 05, France}
\author{Giulia Fischetti}%
\affiliation{Universit\`a di Roma La Sapienza, Rome, Italy}
\author{Enzo Marinari}%
\affiliation{Universit\`a di Roma La Sapienza, Rome, Italy}
\affiliation{CNR-Nanotec and INFN, Sezione di Roma 1, Rome, Italy}
\author{Gleb Oshanin}
\affiliation{Sorbonne Université, CNRS, Laboratoire de Physique Théorique de la Matière Condensée (UMR 7600), 4 Place Jussieu, F-75252 Paris
Cedex 05, France}

\date{\today}

\begin{abstract}
We study the recognition capabilities of the Hopfield model with auxiliary hidden layers, which emerge naturally upon a Hubbard-Stratonovich transformation. We show that the recognition capabilities of such a model at zero-temperature outperform those of the original Hopfield model, due to a substantial increase of the storage capacity and the lack of a naturally defined basin of attraction. The modified model does not fall abruptly in a regime of complete confusion when memory load exceeds a sharp threshold.
\end{abstract}

\maketitle

\textbf{Introduction -} 
Modeling neural networks as Ising spin systems is a rich and interesting field.  Starting from the seminal paper by Little \cite{littleExistencePersistentStates1974},
it was realized that disordered spin systems can 
store information, working as content-addressable memories.
In this context, the model proposed by Hopfield \cite{HOPFIELD1982} (\cH~from now on) has often served as a reference. Its phase diagram  \cite{AGS1987,vanhemmenNonlinearNeuralNetworks1986,forrestContentaddressabilityLearningNeural1988} contains a \textit{retrieval phase}, where one can use a system composed of $N$ neurons to ``store'' patterns containing $N$ symbols. By storage one means that patterns can be recovered: 
starting from the exact pattern we have stored or from a damaged pattern, where a fraction $\eta$ of the spins do not coincide with the configuration we want to retrieve, we end up close enough to it.
Despite being robust in many respects, this model has one essential shortcoming: if $\alpha\equiv P/N$ is larger than a critical value $\alpha_c\approx 0.138$ it is impossible to store more than $P$ different uncorrelated patterns. When $\alpha > \alpha_c$, every memory is abruptly forgotten, and no pattern can be retrieved.  During the last decades, several approaches have been proposed to remedy this and other related issues (see for example
\cite{decelleAsymptoticAnalysisStochastic2011,gardnerOptimalStorageProperties1988,rollsNeuralNetworksBrain,fusiCascadeModelsSynaptically2005,LimitsMemoryStorage,barraEquivalenceHopfieldNetworks2012,agliariParallelRetrievalCorrelated2013,agliariRetrievalCapabilitiesHierarchical2015,baldassiRoleSynapticStochasticity2018,baldassiShapingLearningLandscape2020,schonsbergEfficiencyLocalLearning2021,parisiMemoryWhichForgets1986, marinariForgettingMemoriesTheir2018,vanhemmenForgetfulMemories1988,vanhemmenIncreasingEfficiencyNeural1990}). Here we propose to tackle this problem from a different perspective.

\textbf{Models and Techniques - }
\cH~\cite{HOPFIELD1982}  describes a system of $N$ binary neurons $\sigma_i=\pm1$, $i\in\{1,...,N\}$, with a long range spin glass like Hamiltonian \cite{mezardSpinGlassTheory1986} 
\begin{equation*}
  H[J, \sigma] \; = \; -1/2
  \sum_{i, j=1}^{N} J_{ij} \sigma_{i} \sigma_{j} \,,\quad J_{ij} \; = \; \frac{1}{N} \sum_{\mu=1}^{P} \xi_{i}^{\mu} \xi_{j}^{\mu}.
\end{equation*}
The quenched coupling matrix $J_{ij}$ is defined according to Hebb's learning rule \cite{hebbOrganizationBehaviorNew1950}, where $\xi_{i}^{\mu}=\pm1$, $\mu\in\{1,...,P\}$, $i\in\{1,...,N\}$ are the $P$ configurations (\textit{patterns}) that we want to be able to retrieve. 
The partition function reads
\begin{equation}
Z = \sum_{\{\sigma\}} \; \exp\biggl\{
\frac{\beta}{2N} \sum_{i, j=1}^{N} \sum_{\mu=1}^{P}
 \xi^{\mu}_{i} \xi^{\mu}_{j} \, \sigma_{i} \sigma_{j}
\biggr\} \,,
\label{eq:H_Z}
\end{equation}
where $\beta$ is the inverse temperature of the system.
Recently \cite{mezardMeanfieldMessagepassingEquations2017}
it has been shown that~\cH~can be thought of as the result of a Hubbard-Stratonovich transformation
\begin{equation}
\label{eq:X_Z}
Z \; = \;  \int_{-\infty}^{+\infty} \ldots \int_{-\infty}^{+\infty} \prod_{\mu=1}^{P} dX_{\mu} \;
\sum_{{\bf \sigma}}  \;
\exp\Bigl\{ - \beta \tilde{H}[\xi, {X}, { \sigma}] \Bigr\} \,,
\end{equation}
where the $X_\mu$ are $P$ Gaussian auxiliary variables and
\begin{equation}
\tilde{H}[\xi, X, \sigma] \; \equiv \; N/2  \, \sum_{\mu=1}^{P} X_{\mu}^{2} \; + \; \sum_{\mu=1}^{P} \sum_{i=1}^{N}  \sigma_{i} \, \xi^{\mu}_{i} X_{\mu}.
\label{eq:H_X}
\end{equation}
Integrating over the $X_{\mu}$ in \cref{eq:X_Z} leads back to \cref{eq:H_Z}, and 
corresponds to assuming complete thermalization of the $X_{\mu}$ variables. In this letter we follow a different strategy: we regard the model defined by \cref{eq:X_Z} as fundamental (X model from now on), considering the continuous auxiliary variables as hidden neurons in our system. The hidden neurons enter the learning dynamics on the same footing as the two-state $\sigma$ variables. At $T=0$ energy barriers can and do break the equivalence among the two models, making joint thermalization of the $X_\mu$ and of the $\sigma_i$ variables impossible. Numerical simulations  convincingly demonstrate that this has drastic consequences on the retrieval properties of the system.

At $T=0$, the recognition process in the X model is led by a steepest descent procedure, with sequential updating. One sweep is composed of two steps. First one fixes all the $X_\mu$ variables to minimize the energy given $\sigma_i$: the optimal value is $X^{\text{opt}}_\mu = -(1/N)\sum_i \sigma_i\xi_i^\mu$. Then one updates all the $\sigma_i$ for fixed $\left\{X_\mu\right\}$. When nothing changes in a full sweep we have reached the fixed point.

We study both the X model and ~\cH~ for different values of $\alpha$, $\eta$ and $N$, with $\alpha=0.05$, 0.08, 0.1-0.18 with steps of 0.01, 0.2, 0.22, 0.25, 0.3 (for the X model we have also added simulations at higher $\alpha$ values, both below and above one). We have used $\eta=0$, 0.01, 0.025, 0.05, 0.1, 0.15, 0.2, 0.25 and 0.35. 
For both systems we have studied $10^5$ samples for $N=128$ and $256$, $2\;10^4$ samples for $N=512$, $10^3$ samples for $N=1024$, $10^2$ samples for $N=2048$, $20$ samples for $N=4096$, and a small variable number of samples (normally of order $10$) for $N=8192$.

\textbf{Finite $T$ Monte Carlo - } We first analyzed
the finite temperature structure of both the~\cH~and the
X-model, and verified that at $T>0$ they give the same results. We have implemented an annealing protocol to make this in an effective and controlled way.

\textbf{The average overlap $\aveomega$ - }
\begin{figure}
    \centering
    \includegraphics[width=0.48\textwidth]{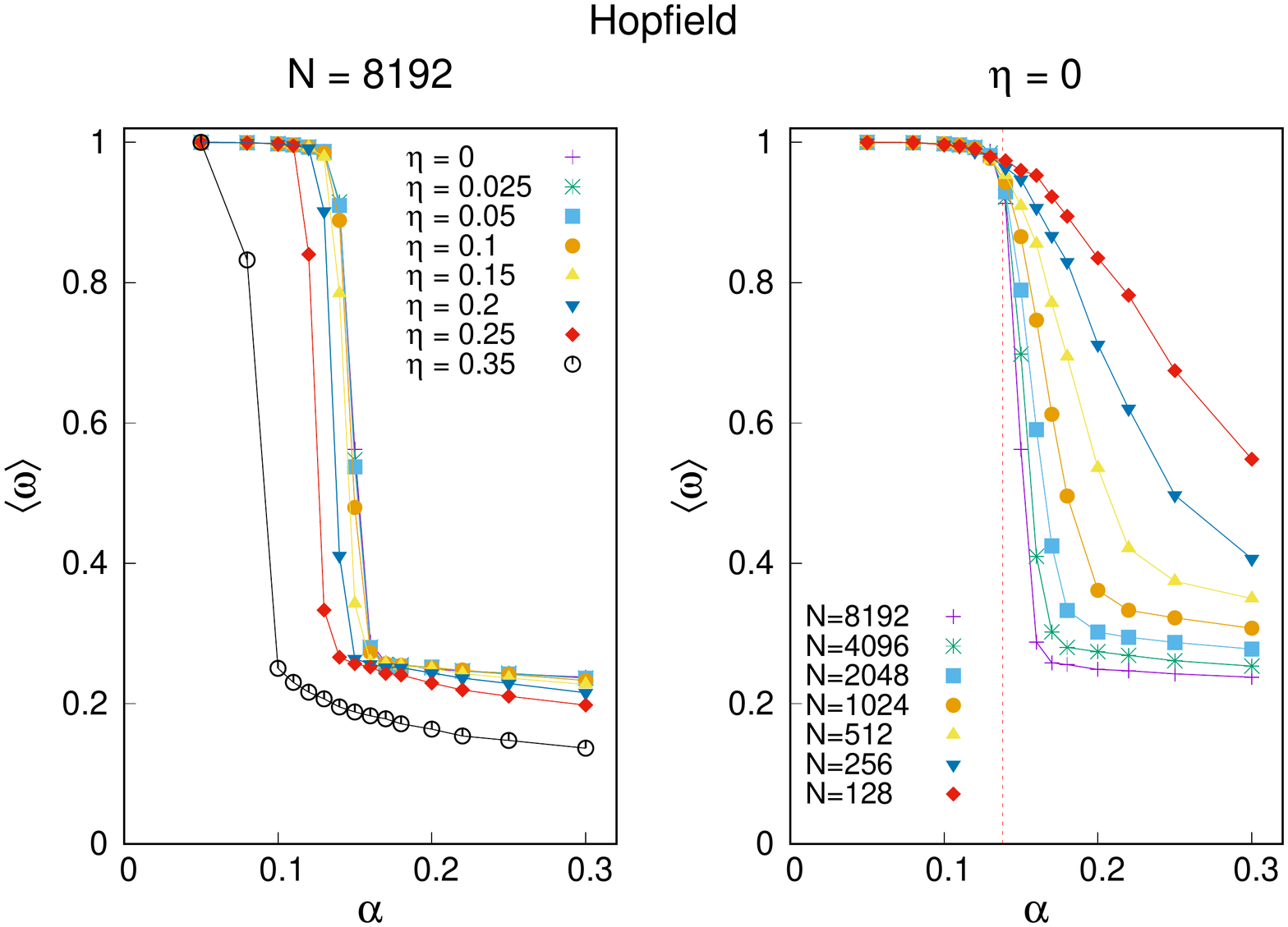}
    \caption{The average overlap $\aveomega$ as a function of $\alpha$ for~\cH.}
    \label{fig:fig1}
\end{figure}
Next we work at $T=0$, with the steepest descent procedure described above. Here the X-model is allowed to behave differently from~\cH. 
We start by measuring the average overlap $\aveomega$  between the starting spin configuration and the stable one where the energy minimization ends (we expect to have recognition when $\aveomega$ is large).
Our results for~\cH~are shown in Fig.~\ref{fig:fig1}. On the left, we use the largest available value of $N$ and plot the values of $\aveomega$ for different values of $\eta$. On the right, we select $\eta=0$ and show the $N$-dependence of $\aveomega$. 
Things for~\cH~work as expected (see \cite{AGS1987}). Increasing $N$ at $\eta=0$ the well known transition forms close to $\alpha\sim 0.136$, where a vertical dashed line is drawn. Upon increasing $\eta$, the transition moves to lower values of $\alpha$, but stays very similar in nature and shape. Even at our highest value of $\eta=0.35$ (where the overlap of the starting point with the original pattern is as low as 0.3), at low $\alpha\lesssim 0.05$ the system is still in recognition phase. Having accurate data for large systems, we are also able to use finite size scaling for a quantitative analysis of these effects (these precision measurements could not fit in this letter, and will be reported in a subsequent publication \cite{BDFMO_2}).

\begin{figure}
    \centering
    \includegraphics[width=0.48\textwidth]{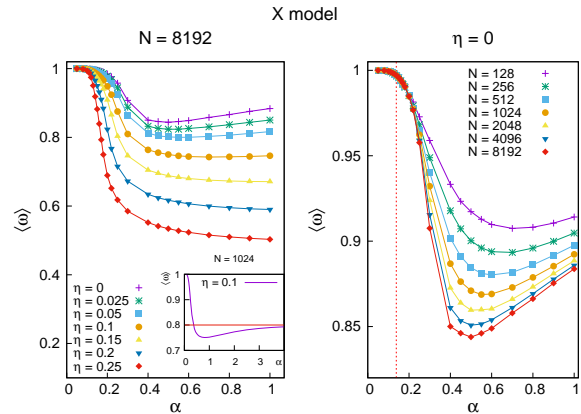}
    \caption{As in Fig. \ref{fig:fig1}, but for the X-model.}
    \label{fig:fig2}
\end{figure}

As shown in Fig.~\ref{fig:fig2} the behavior of the X model is very different. In the $\eta=0$ case the sharp transition of~\cH~contrasts with a smooth, non monotonous behavior of the X-model. At low $\alpha$ we still get a recognition phase, which persists to higher values of $\alpha$ than in~\cH~(and we will attribute to this effect some importance). Increasing $\alpha$ we see that $\aveomega$ smoothly decreases, and reaches a minimum close to $\alpha=0.05$. Here, for $N=8192$ we have $\aveomega\sim 0.84$, but finite size effects are strong. Increasing $\alpha$ further, $\aveomega$ starts to grow. 

This second, high-$\alpha$, regime does not correspond to a recognition phase. To gain some intuition of this, notice that the number of hidden neurons $X_\mu$ in our model is equal to $P$. Hence, it is clear that for very large $\alpha$ the $X_\mu$ are numerous enough to satisfy, by themselves, all the constraints of the problem. In turn, this implies that, as $\alpha\to\infty$, any configuration $\sigma$ can be accommodated in an energy minimum simply by relaxing the hidden neurons to their optimal value, making the dynamic ineffective.
The behavior of the system when $\eta>0$ is very telling. As in~\cH, we still have a recognition region at low $\alpha$, which shrinks for increasing $\eta$. We still have a smooth decrease of $\aveomega$ for increasing $\alpha$, and an asymptotic slow increase that slows down for increasing $\eta$. The asymptotic value for $\alpha\to\infty$ is exactly the initial overlap $1-2\eta$ (inset of Fig.~\ref{fig:fig2}). This clearly confirms that the large $\alpha$ regime is not a recognition regime, but rather a regime of ineffective dynamics. 
 To get a deeper insight about this asymptotic behaviour, we can use the same technique adopted in \cite{FOLERU2017}, and look into what happens under a step of the steepest descent dynamics. Consider any binary neuron configuration $\sigma$. By plugging the expression for $X^{\text{opt}}_\mu$ into the Hamiltonian \cref{eq:H_X}, one sees that the change in energy upon flipping $\sigma_j$, after the $X_\mu$ variables have thermalized to their optimal value given $\sigma$ and the patterns $\xi^\mu$, is
\begin{equation}
\Delta_j\tilde{H}=\frac{2}{N}\sum\limits_{\mu=1}^{P}\Big(1+\sigma_j\xi^\mu_j\sum\limits_{h\neq j}\sigma_h\xi^\mu_h\Big) \,.
\label{eq:deltaH}
\end{equation}
The second contribution in the brackets is what one gets for~\cH~(it shows that $\sigma_j$ is pulled by all the memories $\xi^\mu$, with a strength proportional to the overlap between the memory $\xi^\mu$ and $\sigma$), while the additional $1$ comes because of the $X$ variables interaction. It is a memory independent constant price that one has to pay, due to the fact that we are flipping a spin ``against the will'' of the $X_{\mu}$, which were optimal for $\sigma$. This effectively introduces a threshold in the dynamics of $\sigma$, and a stabilizing effect for the configuration of the binary neurons. This stabilizing effect dominates the dynamics in the high $\alpha$ region, making every configuration stable, and the network useless. On the contrary, and we take this as one of our important findings, the presence of the $X_\mu$ degrees of freedom helps the learning in the small $\alpha$ regime, enlarging\footnote{If we include in the counting the $X_\mu$ degrees of freedom and we redefine $\alpha$ for the X-model accordingly, the comparison is still in favor of the X-model for $\eta=0$, while at higher $\eta$ the two pictures become very similar.} the recognition region as compared to~\cH, and, what is maybe even more important, eliminating complete confusion for $\alpha$ larger than a sharp threshold. 

\begin{figure}
    \centering
    \includegraphics[width=0.48\textwidth]{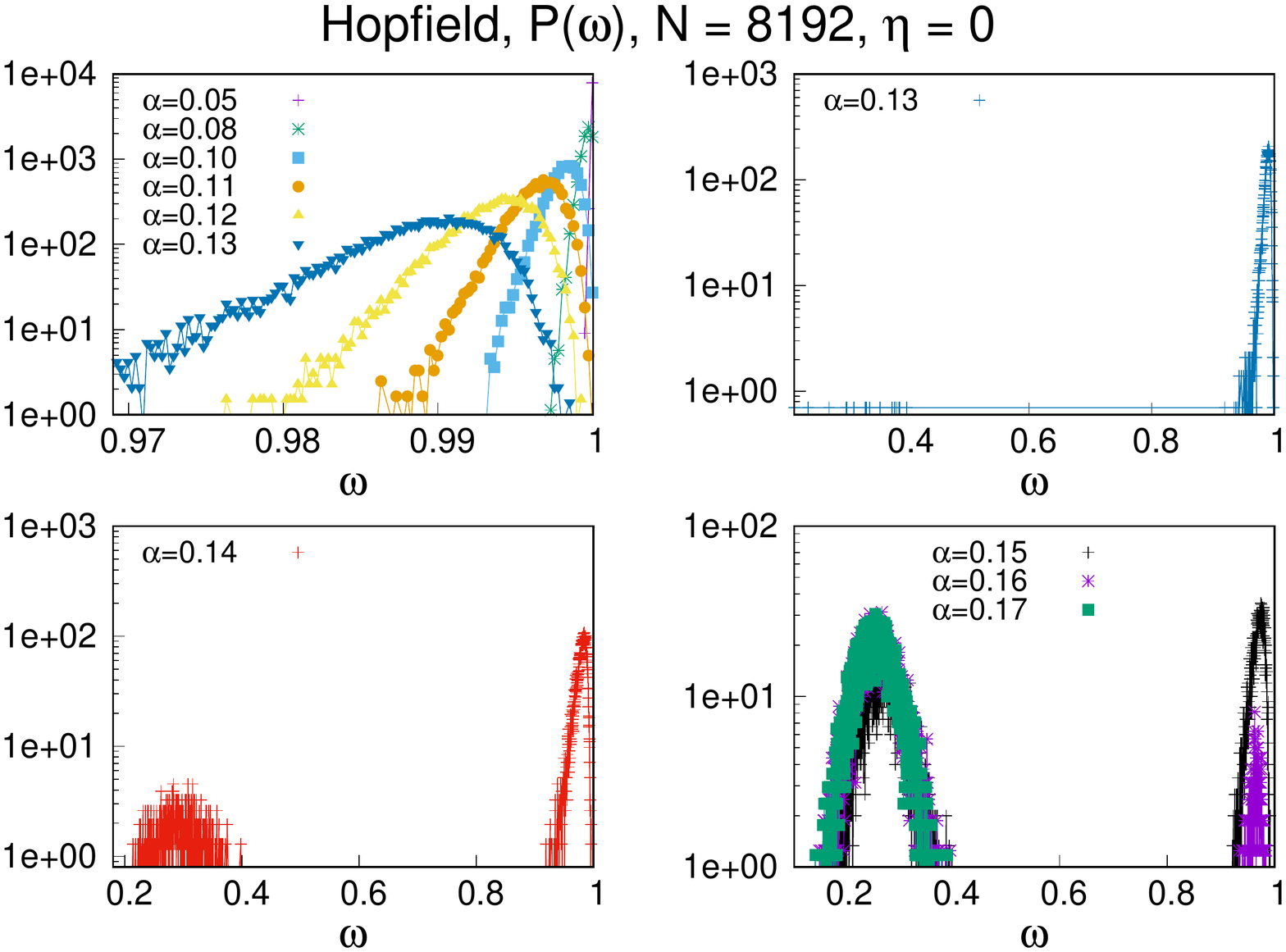}
    \caption{The probability distribution $\pomega$ for ~\cH: $\eta=0$.}
    \label{fig:fig3}
\end{figure}

\textbf{The probability distribution $P(\omega)$ -}
Even if $\aveomega$ is giving us a good amount of information, it is appropriate to analyze the behavior of the full probability distribution $\pomega$. In Fig.~\ref{fig:fig3} we show $\pomega$ at $\eta=0$ for different values of $\alpha$, for~\cH. From left to right and from top to bottom we plot $\pomega$ for increasing values of $\alpha$. The horizontal scales of the four frames are very different. Plots are in linear-log scale. We first show (top left)  results for the low values of $\alpha$, where $\pomega$ is concentrated close to $\omega=1$. Going right from there we plot again $\alpha=0.13$, to show that, just below the critical point, a few points are already at low overlap. The number of these points decreases as $N$ increases.  In the bottom left frame we have $\alpha=0.14$ where, as expected, we have a bi-modality. The peak close to 1 is, at this value of $N$, still leading (remember that the y-scale is a log scale), but a peak at $\omega\sim 0.30$ has appeared. On the right we have the high values of $\alpha$. Now the low $\omega$ peaks start to dominate. Their location is very stable, and only shifts very lightly.  
\begin{figure}
    \centering
    \includegraphics[width=0.48\textwidth]{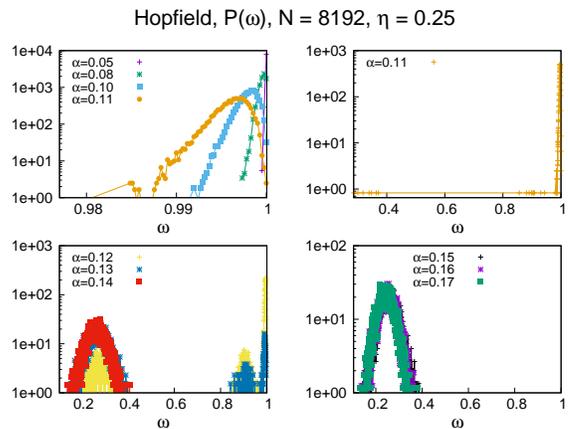}
    \caption{As in Fig. \ref{fig:fig3}, but for $\eta=0.25$.}
    \label{fig:fig4}
\end{figure}

In Fig.~\ref{fig:fig4} we analyze the case $\eta=0.25$ for~\cH. After a rescaling of the value of $\alpha$ everything is analogous to $\eta=0$. The position of the low $\omega$ peak is again remarkably constant. The only clear difference is that here at high-intermediate $\alpha$ value a three peaks structure is visible (there is a clear peak at high $\omega<1$). All together, we find for the structure of~\cH~exactly what we expected.
\begin{figure}
    \centering
    \includegraphics[width=0.48\textwidth]{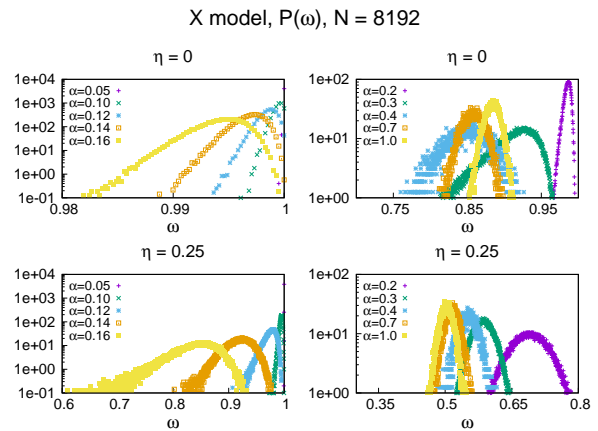}
    \caption{As in Fig. \ref{fig:fig3}, but for the X model: $\eta=0$ and 
    $\eta=0.25$.}
    \label{fig:fig5}
\end{figure}

As we show in Fig.~\ref{fig:fig4}, things are again very different for the X-model. Here we only need two frames for each value of $\eta$ to clearly show  our data. In the X-model we do not see any trace of bi-modality, but only a smooth behavior. For both $\eta=0$ and $\eta=0.25$ at low $\alpha$ we see that the mass of the distribution is centered close to 1. When $\alpha$ increases the distribution first shifts to lower $\alpha$ values, and eventually to larger ones, its $N\to\infty$ limit developing a narrow peak and being centered at $1-2\eta$.

\begin{figure}
    \centering
    \includegraphics[width=0.48\textwidth]{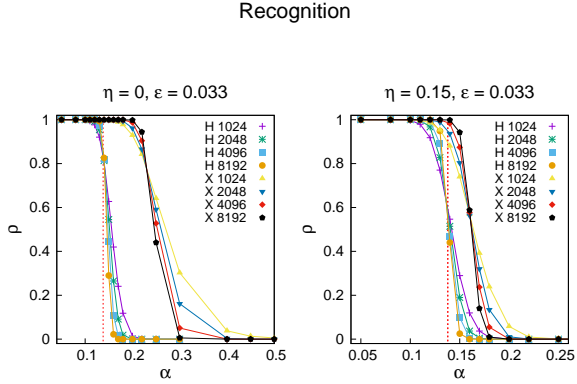}
    \caption{Recognition rate $\rho$ as a function of $\alpha$.}
    \label{fig:fig6}
\end{figure}

\textbf{ The recognition rate - } In order to get further insight, we define the \textit{recognition rate} $\rho$ as the probability that a minimization run ends with $\omega\ge 0.967$ \cite{AGS1987}: the threshold for recognition is $\epsilon=1-0.967=0.033$. In~\cH, that undergoes a sharp transition, selecting a different threshold would give the same asymptotic result. We show in Fig.~\ref{fig:fig6} $\rho$ as a function of $\alpha$, both for~\cH~and the X-model, for $\eta=0$ in the left frame and for $\eta=0.15$ in the right one. Even if, as we have seen in detail, the two models work very differently, the plots for~\cH~and for the X-model are similar, both at $\eta=0$ and at $\eta>0$. The X-model has a wider learning phase. We can say that there is a very low $\alpha$ regime where the new $X$ variables are irrelevant since they are not needed for recognition, and a very large $\alpha$ regime where they fix the system on the observed pattern, but cannot lead to recognition. Only in the region where $\alpha$ is slightly larger than $\alpha_c$, they are put at good use, and help in the memorizing. Also, when $\alpha$ increases, they avoid complete confusion: the X memory becomes less efficient if too many patterns are shown, but the \textit{blackout catastrophe} is avoided.

\textbf{ The dynamical exponent - }
We have also analyzed the rate of the learning dynamics. We assume that the number of sweeps $\mathcal{S}$ needed to reach the stable state scales asymptotically as $N^\zeta$, plus $N$ dependent sub-leading corrections. In a sweep we include, for the X-model, both the cost of putting the $X_\mu$ in their optimal position and the cost of updating every $\sigma_i$ once. 
In absence of any slowing down we expect to find $\zeta=0$.
We define an effective exponent dependent on two values of $N$ as
\begin{equation}
\zeta(N_1,N_2) \equiv \left.\log\left(\mathcal{S}^{(N_1)}\middle/\mathcal{S}^{(N_2)}\right)\middle/\log\left(N_1\middle/N_2\right)\right.\;.
\end{equation}
\begin{figure}
    \centering
    \includegraphics[width=0.48\textwidth]{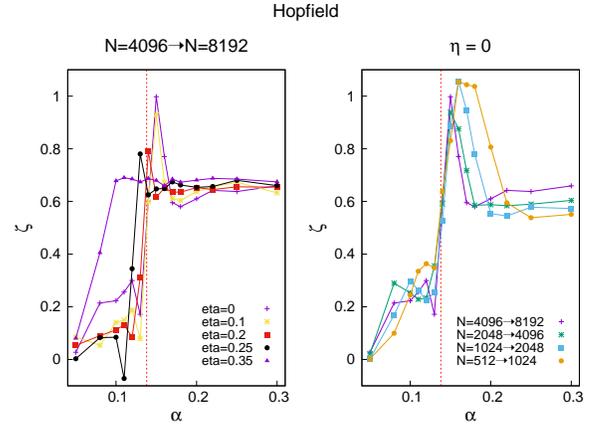}
    \caption{The effective exponent $\zeta(N_1,N_2)$ for~\cH.}
    \label{fig:fig7}
\end{figure}
In Fig.~\ref{fig:fig7} we show $\zeta$ as a function of $\alpha$ for~\cH. On the right $\eta=0$ and different couples of $N$, on the left different $\eta$ values, using $N_1=4096$ and $N_2=8192$. The effective exponent for $\eta=0$ is small at small $\alpha$, is developing a $\delta$-function like peak of value close to one at $\alpha_c$, and is eventually decreasing to an asymptotic large $\alpha$ value close to 0.6. As expected~\cH~is critical close to $\alpha_c$. The situation at $\eta>0$ is similar, but that at $\eta=0.35$ where a sizeable critical peak cannot be detected anymore.

\begin{figure}
    \centering
    \includegraphics[width=0.48\textwidth]{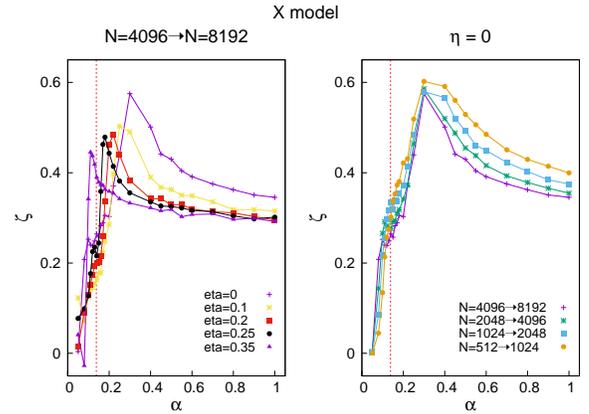}
    \caption{As in Fig. \ref{fig:fig7}, but for the X-model}
    \label{fig:fig8}
\end{figure}
In Fig.~\ref{fig:fig8} we have the same plot for the X-model, and, again, here the situation is different. There is always a peak at low $\alpha$ (larger the $\alpha_c$) but the $N$ dependence is not abrupt, and does not suggest that a $\delta$-function behavior is emerging (even if one would need very large values of $N$ to make sharp claims about this). Also the peak at $\eta>0$ is very different from~\cH, and the effective exponent has a smooth slow decay for large $\alpha$.


\textbf{ Conclusions - } The introduction of hidden layers in the Hopfield model leads to interesting new features in the zero temperature associative memory performance. In our model, the probability distribution of the overlap as well as its average value differ markedly from the ones in the Hopfield model. As a consequence, the recognition performance is improved. More importantly, the interaction between visible and hidden neurons has a stabilizing effect on the zero temperature dynamics, which prevents the \textit{blackout catastrophe}. This, together with the smaller value of the dynamical scaling exponent, implying a faster recognition process, 
suggests that our atypical hidden layers may considerably improve the functioning of Hopfield-like neural systems. This opens an interesting perspective for the further research in the field of associative memory.  

\textbf{Acknowledgments - }We are very grateful to Stefano Fusi and Marc M\'ezard for precious conversations. We have been supported by funding from the European Research Council (ERC) under the European Union's Horizon 2020 research and innovation program (Grant No. 694925-LotglasSy).
\clearpage
\bibliographystyle{apsrev4-1}
\bibliography{maintext.bib}


\end{document}